\documentclass[a4paper,11pt]{article}
\usepackage{pos}
\usepackage{amsmath}
\usepackage{amsfonts}
\usepackage{amsthm}
\usepackage{enumitem}
\usepackage{dcolumn}
\usepackage{bm}
\usepackage{bbm}
\usepackage{cancel}
\usepackage{tikz}
\usetikzlibrary{calc,shapes,arrows,patterns,decorations,positioning,automata,decorations.markings}
\usepackage{tabularx}
\usepackage{booktabs}
\usepackage{tensor}
\usepackage{scalerel}
\usepackage{xcolor}
\usepackage{mathtools}
\usepackage{float}
\usepackage{xfrac}
\definecolor{darkblue}{rgb}{0,0,.5}
\definecolor{darkgreen}{rgb}{0,0.5,.5}
\definecolor{darkyellow}{rgb}{0.5,0.5,0}
\definecolor{fhl}{rgb}{1,0,0}
\usepackage[export]{adjustbox}
\usepackage[most]{tcolorbox}
\tcbuselibrary{skins}
\usepackage{textpos}
\hypersetup{colorlinks=true, breaklinks=true, linkcolor=darkblue, menucolor=darkblue, urlcolor=darkblue, linktocpage=true}

\let\originalleft\left
\let\originalright\right
\renewcommand{\left}{\mathopen{}\mathclose\bgroup\originalleft}
\renewcommand{\right}{\aftergroup\egroup\originalright}

\newcommand{\e}{\operatorname{e}}

\newcommand{\SU}[1]{\operatorname{SU}(#1)}

\newcommand{\of}[1]{\left(#1\right)}

\newcommand{\sbof}[1]{\Bigl(\Big.#1\Big.\Bigr)}
\newcommand{\sof}[1]{\bigl(\big.#1\big.\bigr)}
\newcommand{\ssof}[1]{(#1)}
\newcommand{\fof}[1]{\left[#1\right]}

\newcommand{\cof}[1]{\left\{#1\right\}}

\newcommand{\sscof}[1]{\{#1\}}

\newcommand{\avof}[1]{\left\langle #1\right\rangle}

\newcommand{\trace}{\operatorname{tr}}
\newcommand{\re}{\operatorname{Re}}

\newcommand{\bra}[1]{\left\langle #1\right|}
\newcommand{\ket}[1]{\left| #1\right\rangle}

\newcommand{\dd}{\mathrm{d}}
\newcommand{\DD}[1]{\mathcal{D}\bigl[#1\bigr]}

\newcommand{\partd}[2]{\frac{\partial #1}{\partial #2}}

\newcommand{\abs}[1]{\left| #1\right|}
\newcommand{\ssabs}[1]{| #1|}

\renewcommand*\[{\begin{equation}}
\renewcommand*\]{\end{equation}}
\renewcommand*\hat[1]{\widehat{#1}}
\let\oldstackrel\stackrel
\renewcommand*\stackrel[2]{{\scriptstyle\oldstackrel{#1}{#2}}}

\definecolor{emphcol}{RGB}{0,0,0}
\let\oldemph\emph
\renewcommand*\emph[1]{\oldemph{\textcolor{emphcol}{#1}}}
\let\oldstackrel\stackrel
\renewcommand*\stackrel[2]{{\scriptstyle\oldstackrel{#1}{#2}}}

\newcommand{\ucases}[1]{\begin{cases}#1\end{cases}}

\newcommand*\getscale[1]{%
  \begingroup
    \pgfgettransformentries{\scaleA}{\scaleB}{\scaleC}{\scaleD}{\whatevs}{\whatevs}%
    \pgfmathsetmacro{#1}{sqrt(abs(\scaleA*\scaleD-\scaleB*\scaleC))}%
    \expandafter
  \endgroup
  \expandafter\edef\expandafter#1\expandafter{#1}%
}

\tikzset{fontscale/.style={font=\relsize{#1}}}
\tikzset{->-/.style={decoration={
  markings,
  mark=at position #1 with {\arrow{>}}},postaction={decorate}}}
\tikzset{-<-/.style={decoration={
  markings,
  mark=at position #1 with {\arrow{<}}},postaction={decorate}}}

\tikzset{fontscale/.style={font=\relsize{#1}}}
\tikzset{->-/.style={decoration={
  markings,
  mark=at position #1 with {\arrow{>}}},postaction={decorate}}}
\tikzset{-<-/.style={decoration={
  markings,
  mark=at position #1 with {\arrow{<}}},postaction={decorate}}}

\tikzset{cross/.style={cross out,draw,minimum size=2*(#1-\pgflinewidth),inner sep=0pt, outer sep=0pt}}

\DeclareRobustCommand\sketchdensme{\vcenter{\hbox{\tikz[scale=0.4,nodes={inner sep=0}]{%
\pgfpointtransformed{\pgfpointxy{1}{1}};
  \pgfgetlastxy{\vx}{\vy}
  \def\xs{5};
  \def\ys{4};
  \def\ls{2};
  \begin{scope}[node distance=\vx and \vy]
    \foreach \i in {0,...,\xs} {
        \draw [very thin,gray] (\i,0) -- ($({\i},{0.5*\ys})$)  node[below] at (\i,-0.3) {};
        \draw [very thin,dashed,gray] ($({\i},{0.5*\ys})$) -- ($({\i},{0.5*\ys+1})$);
        \draw [very thin,gray] ($({\i},{0.5*\ys+1})$) -- ($({\i},{\ys+1})$);
    }
    \foreach \i in {0,...,\ys} {
        \draw [very thin,gray] (0,\i) -- (\xs+1,\i) node[left] at (-0.2,\i) {};
    }

    \draw[thick,dashed,red!100] ($({0},{(\ys+1)*1.0+0.025})$) -- ($({(\xs+1)},{(\ys+1)*1.0+0.025})$) node[pos=0.5,anchor=north,yshift=-2pt,scale=1] {$\psi_{2}$};
    \draw[thick,red!100] ($({0},{(\ys+1)*0.0-0.025})$) -- ($({(\xs+1)},{(\ys+1)*0.0-0.025})$) node[pos=0.5,anchor=south,yshift=2pt,scale=1] {$\psi_{1}$};
    
  \end{scope}%
  }}}\,}
  
\DeclareRobustCommand\sketchreddensme{\vcenter{\hbox{\tikz[scale=0.4,nodes={inner sep=0}]{%
  \pgfpointtransformed{\pgfpointxy{1}{1}};
  \pgfgetlastxy{\vx}{\vy}
  \def\xs{5};
  \def\ys{4};
  \def\ls{2};
  \begin{scope}[node distance=\vx and \vy]
    \foreach \i in {0,...,\xs} {
        \draw [very thin,gray] (\i,0) -- ($({\i},{0.5*\ys})$)  node[below] at (\i,-0.3) {};
        \draw [very thin,dashed,gray] ($({\i},{0.5*\ys})$) -- ($({\i},{0.5*\ys+1})$);
        \draw [very thin,gray] ($({\i},{0.5*\ys+1})$) -- ($({\i},{\ys+1})$);
    }
    \foreach \i in {0,...,\ys} {
        \draw [very thin,gray] (0,\i) -- (\xs+1,\i) node[left] at (-0.2,\i) {};
    }
    \draw[thick,dashed,blue!100] ($(0,{(\ys+1)*1.0})$) -- ($({(\xs+1-\ls)},{(\ys+1)*1.0})$) node[pos=0.5,anchor=north,yshift=-2pt,scale=1] {$r_{B}$};
    \draw[thick,blue!100] ($(0,{(\ys+1)*0.0})$) -- ($({(\xs+1-\ls)},{(\ys+1)*0.0})$) node[pos=0.5,anchor=south,yshift=2pt,scale=1] {$r_{B}$};

    \draw[thick,dashed,red!100] ($({(\xs+1-\ls)},{(\ys+1)*1.0+0.025})$) -- ($({(\xs+1)},{(\ys+1)*1.0+0.025})$) node[pos=0.5,anchor=north,yshift=-2pt,scale=1] {$\psi_{A,2}$};
    \draw[thick,red!100] ($({(\xs+1-\ls)},{(\ys+1)*0.0-0.025})$) -- ($({(\xs+1)},{(\ys+1)*0.0-0.025})$) node[pos=0.5,anchor=south,yshift=2pt,scale=1] {$\psi_{A,1}$};
    
    \node[red] at ($({(\xs-0.5*\ls+1)},{(\ys+1)*0.5})$) {$\mathbf{A}$};
    \node[blue] at ($({0.5*(\xs-\ls+1)},{(\ys+1)*0.5})$) {$\mathbf{B}$};
    
  \end{scope}%
  }}}\,}

\title{Improved lattice method for determining entanglement measures in SU(N) gauge theories}
\ShortTitle{Determining entanglement measures in SU(N) lattice gauge theories}

\author*[a]{Tobias Rindlisbacher}
\author[b]{Niko Jokela}
\author[c]{Arttu P\"onni}
\author[b]{Kari Rummukainen}
\author[b]{Ahmed Salami}

\affiliation[a]{AEC, Institute for Theoretical Physics, University of Bern, Sidlerstrasse 5, CH-3012 Bern, Switzerland}
\affiliation[b]{Helsinki Institute of Physics and Department of Physics, P.O. Box 64, FI-00014 University of Helsinki, Finland}
\affiliation[c]{Micro and Quantum Systems Group, Department of Electronics and Nanoengineering, Aalto University, Finland}

\emailAdd{trindlis@itp.unibe.ch}

\abstract{The determination of entanglement measures in SU(N) gauge theories is a non-trivial task. With the so-called "replica trick", a family of entanglement measures, known as "Rényi entropies", can be determined with lattice Monte Carlo. Unfortunately, the standard implementation of the replica method for SU(N) lattice gauge theories suffers from a severe signal-to-noise ratio problem, rendering high-precision studies of Rényi entropies prohibitively expensive.
In this work, we propose a method to overcome the signal-to-noise ratio problem and show some first results for SU(N) in 4 dimensions.
\vskip .1cm
{\footnotesize  \it Preprint:  HIP-2022-26/TH}}

\FullConference{%
The 39th International Symposium on Lattice Field Theory,\\
8th-13th August, 2022,\\
Rheinische Friedrich-Wilhelms-Universität Bonn, Bonn, Germany
}

\begin{document}
\maketitle
\section{Introduction}\label{sec:intro}
Consider the Hilbert space of an isolated quantum system consisting of two disjoint parts, $A$ and $B$, i.e. $\mathcal{H}_{AB}=\mathcal{H}_A\otimes\mathcal{H}_B$. Each state $\ket{\psi}_{AB}\in\mathcal{H}_{AB}$ can be described in terms of orthonormal bases $\ket{n}_{A}\in\mathcal{H}_{A}$ and $\ket{m}_{B}\in\mathcal{H}_{B}$ of the subsystems as
\[
\ket{\psi}_{AB}=\sum_{m n} a_{m n}\,\ket{m}_A\otimes\ket{n}_B\quad,\quad \sum_{m n}\abs{a_{m n}}^2=1\ ,\label{eq:purestateAB}
\]
with corresponding density operator\footnote{We use notation where the tensor product between vectors from dual Hilbert spaces, e.g. between $\ket{\psi}_A\in\mathcal{H}_{A}$ and ${}_A\bra{\phi}\in\mathcal{H}^{*}_{A}$, is written as $\ket{\psi}_A\bra{\phi}\in\mathcal{H}_{A}\otimes\mathcal{H}^{*}_{A}$. All other tensor products are written with explicit $\otimes$-symbols, as e.g. $\ket{\psi}_A\otimes{}\ket{\phi}_A\in\mathcal{H}_{A}\otimes\mathcal{H}_{A}$, $\ket{\psi}_A\otimes{}\ket{\phi}_B\in\mathcal{H}_{A}\otimes\mathcal{H}_{B}$ or $\ket{\psi}_A\otimes{}_B\bra{\phi}\in\mathcal{H}_{A}\otimes\mathcal{H}^{*}_{B}$.}
\[
\rho_{AB}=\ket{\psi}_{AB}\bra{\psi}=\sum_{m n k l} a^{\phantom{*}}_{m n} a^{*}_{k l} \ket{m}_A\bra{k}\otimes\ket{n}_B\bra{l}\ .\label{eq:purestaterhoAB}
\]
As this density operator corresponds to a \emph{pure state}, it has the projector property,
\[
\rho^2=\rho\ .\label{eq:projectproppurerho}
\]
General density operators, however, can be convex sums of pure states (not necessarily orthogonal),
\[
\rho=\sum_{i}\,p_i\,\ket{\psi_i}\bra{\psi_i}\quad,\quad \ket{\psi_i}\in\mathcal{H}_{AB}\,\forall\,i\quad,\quad \sum_{i}\,p_i=1\ ,\label{eq:genrho}
\]
and do not have the projection property Eq.~\eqref{eq:projectproppurerho}. For such \emph{mixed states} one always has $\trace\ssof{\rho^2}<1$.

From the pure state density operator in Eq.~\eqref{eq:purestaterhoAB} we can trace out e.g. subsystem $B$ to obtain a reduced density matrix for subsystem $A$:
\[
\rho_{A}=\trace_{B}\of{\rho_{AB}}=\sum_{m k l} a^{\phantom{*}}_{m l} a^{*}_{k l} \ket{m}_A\bra{k}\ .\label{eq:reducedrhoA}
\]
The reduced density matrix in Eq.~\eqref{eq:reducedrhoA} is in general a mixed state, unless
\[
a_{mn}=a^{\ssof{A}}_{m} a^{\ssof{B}}_{n}\quad\forall\,m,n\quad,\quad \sum_{m} \ssabs{a^{\ssof{A}}_{m}}^2=1\quad,\quad \sum_{n} \ssabs{a^{\ssof{B}}_{n}}^2=1\ ,
\]
i.e. unless $\rho_{AB}$ is a product sate, $\rho_{AB}=\rho_{A}\otimes\rho_{B}$. If the total system is in such a product state, then the subsystems $A$ and $B$ are completely independent and not entangled. If, however the system is described by a general pure state, Eq.~\eqref{eq:purestaterhoAB}, which is not a product state, so that Eq.~\eqref{eq:reducedrhoA} satisfies $\trace\ssof{\rho^2_{A}}<1$, then the subsystems $A$ and $B$ are entangled. This motivates the definition of a family of measures for quantifying the entanglement of a subsystem $A$ with its complement $B$, based on how quickly $\trace\ssof{\rho_A^s}$ decays with increasing $s$, as for example the \emph{R{\'e}nyi entropies} of order $s\in\cof{2,3,\ldots}$~\cite{Renyi:1965aa},
\[
H_{s}\of{A}=\frac{1}{1-s}\log\trace\of{\rho_{A}^{s}}\ ,\label{eq:rsentropy}
\] 
or the \emph{entanglement entropy},
\[
S_{EE}\of{A}=-\lim_{s\to 1} \partd{\log\trace\ssof{\rho_A^s}}{s}=-\lim_{s\to 1} \partd{\of{\of{1-s} H_s\of{A}}}{s}=\lim_{s\to 1} H_s\of{A}\ .\label{eq:eentropy}
\]
The latter can with the spectral theorem also be written as Von Neumann entropy of the reduced density matrix $\rho_A$:
\[
S_{EE}\of{A}=-\trace\of{\rho_A\log\of{\rho_A}}\ .\label{eq:eentropydef}
\]

If $S_{EE}\of{A}=0$ then subsystems $A$ and $B$ are not entangled. The larger the value of $S_{EE}\of{A}$, the more mixed $\rho_A$ is and the more entangled $A$ and $B$ are. The same holds when using $H_2\of{A}$ instead of $S_{EE}\of{A}$. However, this discussion applies only if the whole system is described by a pure state $\rho_{AB}$. If one has a finite temperature partition function at temperature $T$:
\[
Z_{AB}\of{\beta}=\trace\of{\rho_{AB}}\quad,\quad \rho_{AB}=\sum_{n} p_n\,\ket{\psi_n}\bra{\psi_n}\quad,\quad \ket{\psi_n}\in\mathcal{H}_{AB}\quad,\quad p_n=\frac{\e^{-E_n/T}}{\sum_{i}\e^{-E_i/T}}
\]
where $\sscof{\ssof{E_n,\ket{\psi_n}}}_{n=0,1,\ldots}$ are the energy eigenvalues and corresponding eigenstates for the full system, then the system is obviously described by a mixed state, and only in the zero-temperature limit, $\of{T\to 0}$, $\rho_{AB}$ reduces to a pure state, namely the ground state. We will not try to quantify entanglement for mixed states. Results obtained with the methods described in the remainder of this article should therefore be considered as lattice approximations to the zero-temperature entanglement and R{\'e}nyi entropies (resp. their derivatives). 

\section{Replica method on the lattice}\label{sec:replicamethod}
Consider a $\SU{N}$ gauge theory on a $d$-dimensional, finite, Euclidean lattice of temporal extent $N_t$. We can compute the matrix elements of a corresponding density matrix, $\rho$, using a Euclidean path-integral representation:
\[
\bra{\psi_{1}}\rho\ket{\psi_{2}}=\frac{1}{Z}\int\limits_{\mathclap{\substack{U\ssof{\bar{x},0}=U_{\psi_1}\ssof{\bar{x}}\\ U\ssof{\bar{x},N_t}=U_{\psi_2}\ssof{\bar{x}}}}}\DD{U}\,\e^{-S_G\fof{U}}=\frac{1}{Z}\sketchdensme\ ,\label{eq:densmatpirep}
\]
where the states $\psi_1$ and $\psi_2$ define (up to gauge) values for the gauge links that touch the time slices $x^d=0$ and $x^d=N_t$, respectively. The normalization factor $Z$ is obtained from the normalization condition
\[
1=\trace\of{\rho}=\int\dd{\psi}\bra{\psi}\rho\ket{\psi}=\frac{1}{Z}\int\DD{U}\,\e^{-S_{G}\fof{U}}\ ,\label{eq:densmattrace}
\]
and is just the usual Euclidean lattice partition function with periodic boundary conditions:
\[
Z=\int\DD{U}\,\e^{-S_{G}\fof{U}}\ .\label{eq:ordinpartf}
\]

\begin{figure}[h]
\centering
\begin{minipage}[t]{0.49\linewidth}
\centering
\includegraphics[width=0.75\linewidth]{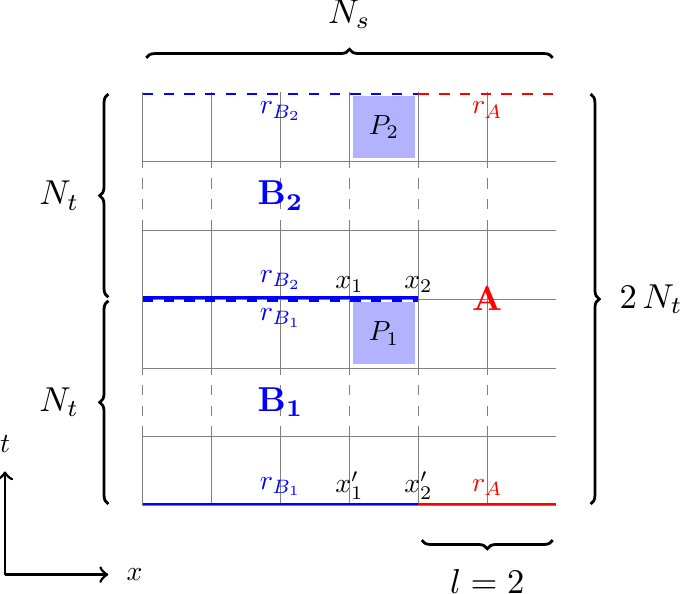}
\end{minipage}\hfill
\begin{minipage}[t]{0.49\linewidth}
\centering
\includegraphics[width=0.75\linewidth]{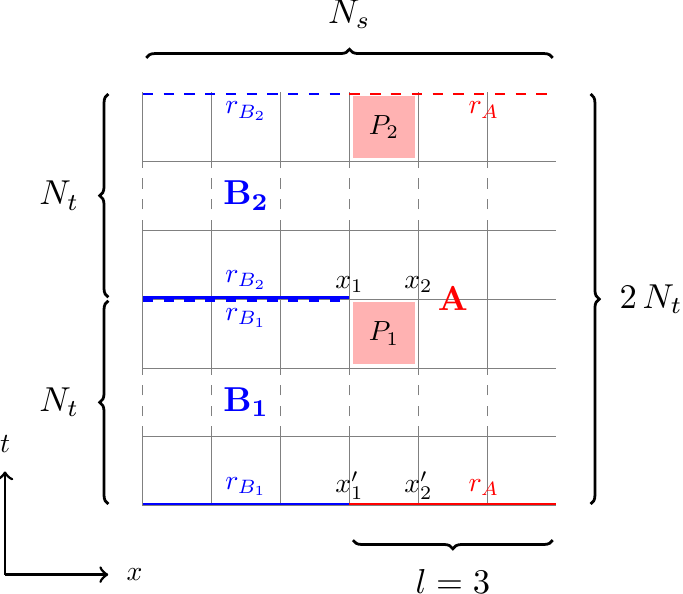}
\end{minipage}\\[5pt]
\begin{minipage}[t]{0.49\linewidth}
\centering
\includegraphics[width=0.75\linewidth]{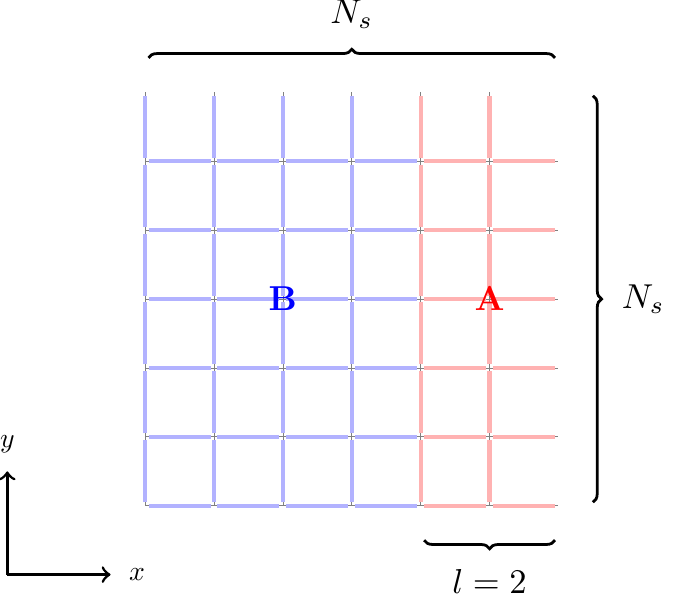}
\end{minipage}\hfill
\begin{minipage}[t]{0.49\linewidth}
\centering
\includegraphics[width=0.75\linewidth]{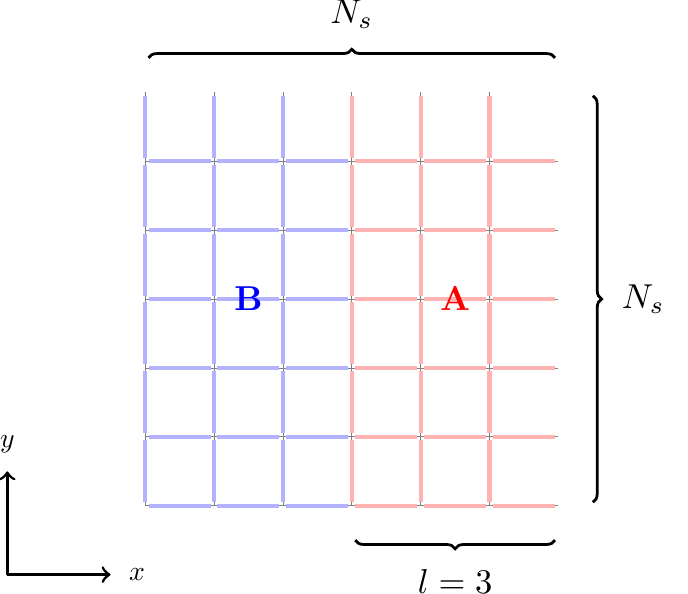}
\end{minipage}
\caption{The figure illustrates the topology of the system described by the partition function $Z_c\of{l,s,N_t,N_s}$ that enters Eq.~\eqref{eq:renyientropyords} at the example of a (2+1)-dimensional lattice of size $N_{s}^2\times s\,N_{t}$ with $s=2$ and $l=2$ (left) resp. $l=3$ (right). The upper panels show the situation in the x-t-plane to visualize the different boundary conditions in regions $A$ and $B$: dashed and solid vertical lines with the same labels $r_{B_1}$, $r_{B_2}$, or $r_A$ are identified so that when approaching a dashed vertical line from below, one arrives at the corresponding solid vertical line. The two lower panels show the situation in the x-y-plane where the spatial links over which the temporal plaquettes are subject to the boundary conditions of region $A$ are marked in red and those over which the temporal boundary conditions of region $B$ apply are marked in blue.\\
A change of boundary conditions over a spatial link affect the way in which temporal plaquettes over that link are computed: the plaquette $P_{1}$ in the top-left panel touches from below the dashed line $r_{B_1}$, so that the top-edge of $P_{1}$ does not consist of the link variable that connects $x_{1}$ and $x_{2}$, but rather the link variable that connects $x'_{1}$ and $x'_{2}$. Similarly, the upper edge of plaquette $P_{2}$ does not consist of the link between $x'_{1}$ and $x'_{2}$ but of the link between $x_{1}$ and $x_{2}$. In the top-right panel, the same temporal plaquettes are marked in red and belong now to region $A$: the link that closes the plaquette $P_{1}$ along its top-edge is now the one that connects $x_{1}$ and $x_{2}$ and the link that closes $P_{2}$ along its top-edge is the one that connects $x'_{1}$ to $x'_{2}$.}
\label{fig:bdcond}
\end{figure}

If the lattice system is split into two parts, $A$ and $B$, as depicted in the left-hand part of Fig.~\ref{fig:bdcond}, the matrix elements of the reduced density matrix for part $A$ is given by
\[
\bra{\psi_{A,1}}\rho_A\ket{\psi_{A,2}}=\int\dd{\psi_B}\bra{\psi_{B}\otimes\psi_{A,1}}\rho\ket{\psi_{B}\otimes\psi_{A,2}}=\frac{1}{Z}\sketchreddensme\ ,\label{eq:reddensmatpirep}
\]
where in contrast to the full trace in Eq.~\eqref{eq:densmattrace} only the temporal boundary states, $\psi_{B,1}$ and $\psi_{B,2}$ for part $B$ get identified with each other and summed. In the sketch after the last equality sign, this is represented by assigning the label $r_B$ to both temporal boundaries of part $B$. The states $\psi_{A,1}$, $\psi_{A,2}$ define only the temporal boundaries for part $A$.

With these definitions it is now straightforward to represent the trace of powers of the reduced density matrix, $\rho_A$, in terms of ratios of two lattice partition functions (cf. replica method~\cite{Calabrese:2004eu}) and write the \emph{R{\'e}nyi entropies} from Eq.~\eqref{eq:rsentropy} for $s\geq 2$ as:
\[
H_{s}\of{l,N_t,N_s}=\frac{1}{1-s}\log\frac{Z_{c}\of{l,s,N_{t},N_{s}}}{Z^{s}\of{N_{t},N_{s}}}=\frac{-1}{1-s}\of{F_c\of{l,s,N_t,N_s}-s\,F\of{N_t,N_s}}\ .\label{eq:renyientropyords}
\]
Here $Z\of{N_t,N_s}$ is the ordinary partition function from Eq.~\eqref{eq:ordinpartf} for a lattice of size $N_s^{d-1}\times N_t$ and $Z_c\of{l,s,N_t,N_s}$ is the partition function for a system with topology as depicted in Fig.~\ref{fig:bdcond}. $F_c\of{l,s,N_t,N_s}$ and $F\of{N_t,N_s}$ are the corresponding free energies. Note that $Z_c\of{l=0,s,N_t,N_s}=Z^s\of{N_t,N_s}$ and $Z_c\of{l,s=1,N_t,N_s}=Z\of{N_t,N_s}$.

The entanglement entropy from Eq.~\eqref{eq:eentropy} cannot be written in closed form in terms of ratios of partition functions. In lattice studies one therefore makes use of the approximation,
\begin{multline}
S_{EE}\of{l,N_t,N_s}=-\lim_{s\to 1} \partd{\log\trace\of{\rho_A^s}}{s}=\lim_{s\to 1}\partd{F_c\of{l,s,N_t,N_s}}{s}-F\of{N_t,N_s}\\
\approx F_c\of{l,2,N_t,N_s}-2\,F\of{N_t,N_s}=H_{2}\of{l,N_t,N_s}\ .\label{eq:eentropylat}
\end{multline}
The accuracy of this approximation is discussed in~\cite{Rabenstein:2018bri}.

\section{Overlap problem}\label{sec:overlapproblem}
When trying to determine the free energy difference in Eq.~\eqref{eq:eentropylat} with Monte Carlo simulations by direct measurements,
\[
H_2\of{l,N_t,N_s}=-\log\frac{Z_c\of{l,2,N_t,N_s}}{Z_c\of{0,2,N_t,N_s}}=-\log\,\avof{\exp\of{-S_{G,l}\fof{U}+S_{G,0}\fof{U}}}_{Z_c\of{0,2,N_t,N_s}}\ ,\label{eq:zratioreweight}
\]
where $S_{G,l}\fof{U}$ is for a given gauge link configuration, $U=\sscof{U_{x,\nu}}_{x,\nu}$, the value of the Wilson gauge action for the lattice topology that corresponds to an entangling region $A$ of width $l$ (cf. Fig.~\ref{fig:bdcond}) and $\avof{\ldots}_{Z}$ is the expectation value with respect to the partition function $Z$,
one will be facing a severe overlap problem: link-variable configurations that (significantly) contribute to the partition function $Z_c\of{l=0,2,N_t,N_s}$ play essentially no role for the partition function $Z_c\of{l,2,N_t,N_s}$ (and vice versa); a reliable estimate on the log of the observable on the right-hand side of Eq.~\eqref{eq:zratioreweight} would therefore require exponentially large statistics.

The overlap problem gets somewhat milder when looking at the derivative
\[
\left.\partd{S_{EE}\of{l',N_t,N_s}}{l'}\right\vert_{l'=l+1/2}\approx F_c\of{l+1,2,N_t,N_s}-F_c\of{l,2,N_t,N_s}\ \label{eq:eentropylatderiv}
\]
instead of at Eq.~\eqref{eq:eentropylat} itself, which is often done also in a non-lattice context in order to get rid of a $l$-independent UV-divergent piece in $S_{EE}\of{l,N_t,N_s}$~\cite{Buividovich:2008kq,Nakagawa:2009jk,Rabenstein:2018bri}. However, also for the free energy difference in Eq.~\eqref{eq:eentropylatderiv} the overlap problem will in general still be too severe to allow for a naive measurement {\`a} la Eq.~\eqref{eq:zratioreweight}.

In previous works~\cite{Buividovich:2008kq,Nakagawa:2009jk,Rabenstein:2018bri} the overlap problem has been addressed by determining Eq.~\eqref{eq:eentropylatderiv} from
\[
\Delta F\of{l}=-\int\limits_0^1\dd\alpha\,\partd{\log Z^{*}_{l}\of{\alpha}}{\alpha}=\int\limits_0^1\dd\alpha\,\avof{S_{G,l+1}-S_{G,l}}_{Z^{*}_{l}\of{\alpha}}\ ,\label{eq:dseedlint}
\]
using an interpolating partition function
\[
Z^{*}_{l}\of{\alpha}=\int\DD{U}\,\exp\of{-\of{1-\alpha}\,S_{G,l}\fof{U}-\alpha\,S_{G,l+1}\fof{U}}\ .\label{eq:partftradip}
\]
This approach suffers unfortunately form a bad signal to noise ratio for the integrated observable, Eq.~\eqref{eq:dseedlint}, which is presumably due to a huge free energy barrier in Eq.~\eqref{eq:partftradip} as $\alpha$ runs from 0 to 1. 

\section{Improved interpolation method}\label{sec:overlapproblemimproved}
In this work, we try to avoid the formation of huge free energy barriers when measuring \eqref{eq:eentropylatderiv} by using a different interpolation method. Let us denote by $C$ the set of all plaquettes in a $V=s\,N_{t}\times N_{s}^{d-1}$ lattice which have the same value regardless of whether their spatial location is part of region $A$ or region $B$. These are all plaquettes but the temporal ones that touch from below the time-slices with $x^{d}=r\cdot N_{t}$ for $r=1,\ldots,s$ (cf. Fig.~\ref{fig:bdcond}). A generalized partition function can then be written as:
\begin{multline}
Z\sof{\beta,s,N_{t},N_{s},\cof{n}}=\\
\int\DD{U}\exp\sbof{\frac{\beta}{N}\sbof{\sum\limits_{\Box \in C}\re\trace\sof{U_{\vphantom{\bar{l}}\Box}}+\sum_{\bar{x}}\sum_{\nu=1}^{d-1}\sum_{r=1}^{s}\re\trace\sof{U^{\of{n_{\bar{x},\nu}}}_{\nu d}\of{\bar{x},r\cdot N_{t}-1}}}}\ ,\label{eq:genpartf}
\end{multline}
where $\bar{x}={x^{1},\ldots,x^{d-1}}\in\mathbb{Z}^{d-1}$ labels the spatial positions on the lattice and $U^{\of{0}}_{\nu d}\of{x}$ and $U^{\of{1}}_{\nu d}\of{x}$ are the two different values for the temporal plaquette
\[U_{\nu d}\of{x}=U^{\vphantom{\dagger}}_{\nu\vphantom{\hat{\nu}}}\of{x}\,U^{\vphantom{\dagger}}_{d}\of{x+\hat{\nu}}\,U^{\dagger}_{\nu}\of{x+\smash{\hat{d}}\vphantom{\hat{\nu}}}\,U^{\dagger}_{d}\of{x\vphantom{\hat{\nu}}}\ ,
\] 
depending on whether the plaquette is subject to the temporal boundary conditions from outside or inside region A (see Fig.~\ref{fig:bdcond}: (0) corresponds to blue, (1) to red). Which case applies for which spatial link is controlled by the set $\cof{n}=\sscof{n_{\bar{x},\nu}}_{\bar{x}\in\mathbb{Z}^{d-1},\nu\in\cof{1,\ldots,d-1}}$ of discrete variables $n_{\bar{x},\nu}\in\cof{0,1}$. 

To describe how the interpolation between $Z_{c}\of{l,s,N_{t},N_{s}}$ and $Z_{c}\of{l+1,s,N_{t},N_{s}}$ can be carried out using the generalized partition function Eq.~\eqref{eq:genpartf}, let us denote by $K=\sscof{\ssof{\bar{x}_{i},\nu_{i}}}_{i=1,\ldots,N_{K}}$
the ordered set of spatial links in the boundary region where $N_{s}-l-1\leq x_{1}<N_{s}-l$, and let $K_{j}=\sscof{\ssof{\bar{x}_{i},\nu_{i}}}_{i=1,\ldots,j}$ be the subset of the first $j$ elements of $K$. Let us then further define:
\[
n^{i}_{\bar{x},\nu}=\ucases{0\quad\text{if}\quad x_{1}<N_{s}-l-1\quad\text{or}\quad \of{\bar{x},\nu}\in K\setminus K_{i}\\ 1\quad\text{if}\quad x_{1}\geq N_{s}-l\quad \text{or}\quad \of{\bar{x},\nu}\in K_{i}}\ ,
\]
and abbreviate $Z_{i}=Z\ssof{\beta,s,N_{t},N_{s},\sscof{n^{i}}}$ and $F_{i}=-\log\ssof{Z_{i}}$.
The expression for the derivative of the entanglement entropy \eqref{eq:eentropylatderiv} then becomes:
\[
\left.\partd{S_{EE}\of{l',N_t,N_s}}{l'}\right\vert_{l'=l+1/2}\approx-\log\ssof{Z_{N_{K}}}+\log\ssof{Z_{0}}=F_{N_{K}}-F_{0}\ .\label{eq:imprfreeenergydiff}
\]
This free energy difference can be measured from a single simulation using  Wang-Landau (WL) sampling~\cite{Wang:2000fzi}. With this method one samples the modified partition function,
\begin{multline}
Z_{\text{WL}}\ssof{\beta,s,N_{t},N_{s},\cof{f}}=\\
\sum\limits_{i=0}^{N_{K}} e^{f_{i}}\int\DD{U}\exp\sbof{\frac{\beta}{N}\sbof{\sum\limits_{\Box \in C}\re\trace\sof{U_{\vphantom{\bar{l}}\Box}}+\sum_{\bar{x}}\sum_{\nu=1}^{d-1}\sum_{r=1}^{s}\re\trace\sof{U^{\ssof{n^{i}_{\bar{x},\nu}}}_{\vphantom{\bar{x}}\smash{\bar{x}+\ssof{r\cdot N_{t}-1}\cdot\hat{d},\nu d}}}}}\ ,\label{eq:wlpartf}
\end{multline}
and adjusts the set of parameters $\cof{f}=\sscof{f_{i}}_{i=0,\ldots,N_{K}}$ till the histogram $\cof{H}=\sscof{H_{j}}_{j=0,\ldots,N_{K}}$, with
\[
H_{j}=\partd{\log\of{Z_{\text{WL}}\sof{\beta,s,N_{t},N_{s},\cof{f}}}}{f_{j}}{}\ ,\label{eq:wlhistentry}
\]
is approximately flat. After appropriate values for $\cof{f}$ have been found with the WL method, one can start to accumulate high statistics for the histogram $\cof{H}$ while keeping the $\cof{f}$ fixed, and obtain an improved estimator for the free energy difference by setting:
\[
F_{N_{K}}-F_{0}=f_{N_{K}}-f_{0}+\log\ssof{H_{N_{K}}}-\log\ssof{H_{0}}\ .\label{eq:imprfreeenergydiff2}
\] 
Error bars for Eq.~\eqref{eq:imprfreeenergydiff2} can be obtained from the histograms in Eq.~\eqref{eq:wlhistentry} using the jackknife method.
\begin{figure}[h]
\centering
\begin{minipage}[t]{0.475\linewidth}
\centering
\includegraphics[width=\linewidth,keepaspectratio,bgcolor=black!2]{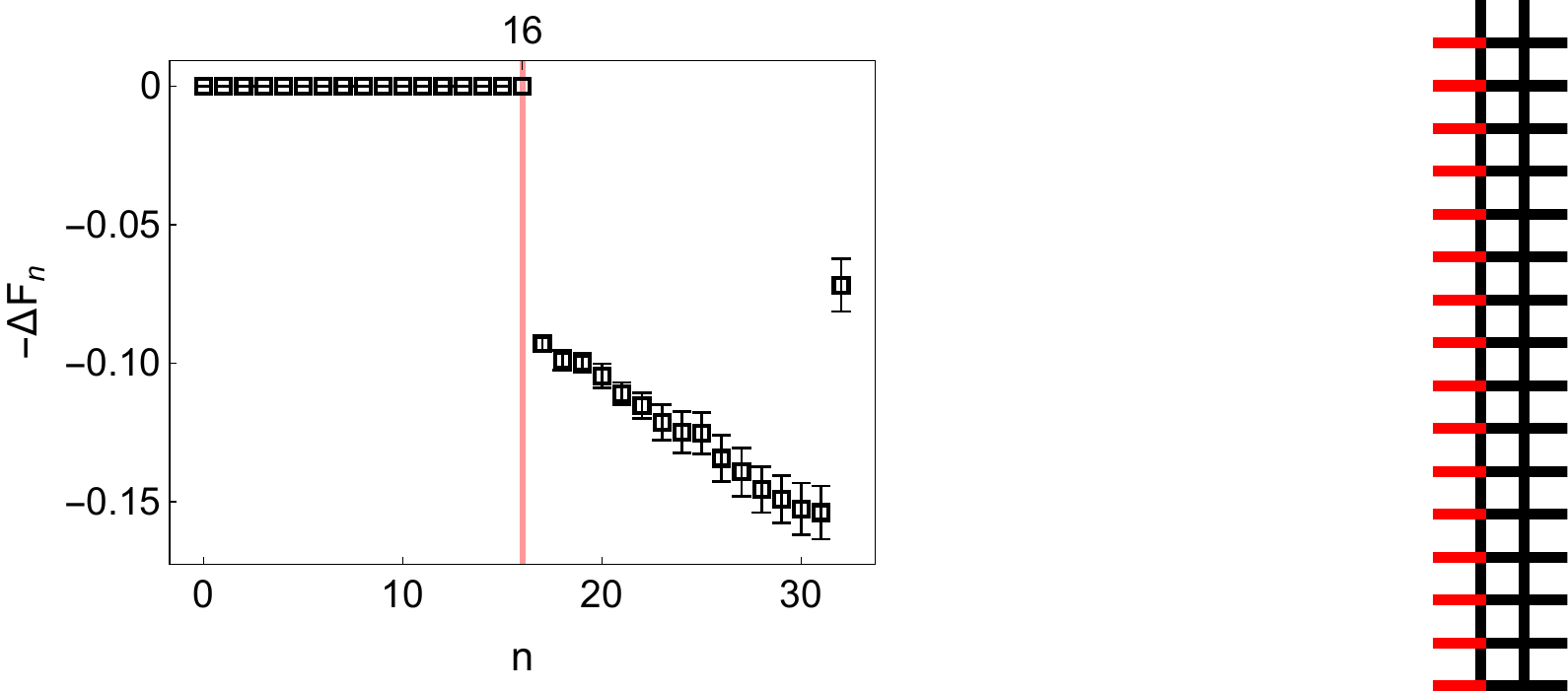}
\end{minipage}\hfill
\begin{minipage}[t]{0.475\linewidth}
\centering
\includegraphics[width=\linewidth,keepaspectratio,bgcolor=black!2]{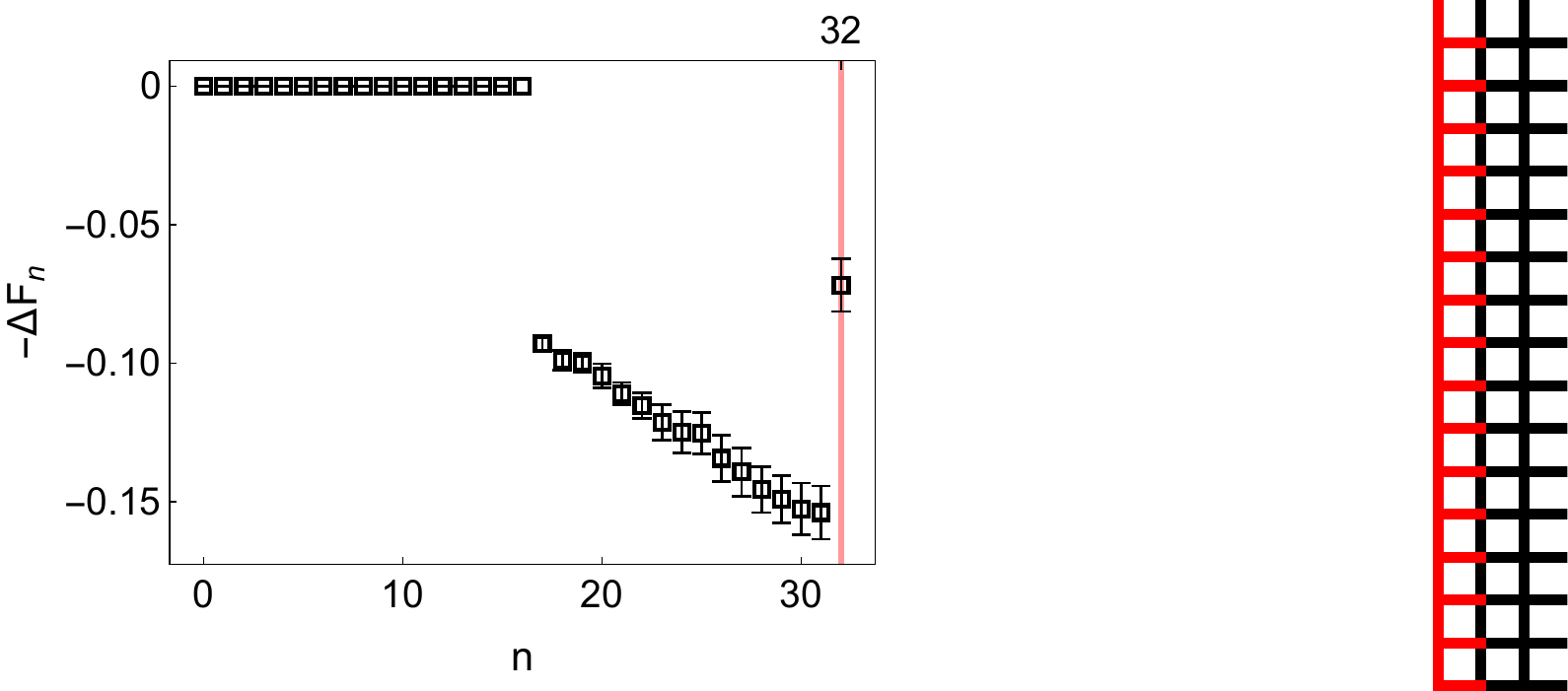}
\end{minipage}
  \caption{Illustration of how in $\SU{2}$ gauge theory on a (2+1)-dimensional lattice with topology as in the left-hand panel of Fig.~\ref{fig:bdcond}, with $N_s=16$, $N_t=16$, $s=2$, and $\beta=5.0$, the free energy changes as individual spatial links are added to region $A$ to grow its width from $l=2$ to $l=3$. The spatial links perpendicular to the boundary of region $A$ are added first, and only afterwards the parallel ones are added. On the left-hand side of each panel, the total change in free energy as function of the number $n$ of spatial links added to region A is shown. The red, vertical lines in these plots mark the $n$-values that corresponds to the sketched boundaries.}
  \label{fig:freeenergyvsn3d}
\end{figure}
\begin{figure}[h]
\centering
\begin{minipage}[t]{0.495\linewidth}
\centering
\includegraphics[width=\linewidth,keepaspectratio,bgcolor=black!2]{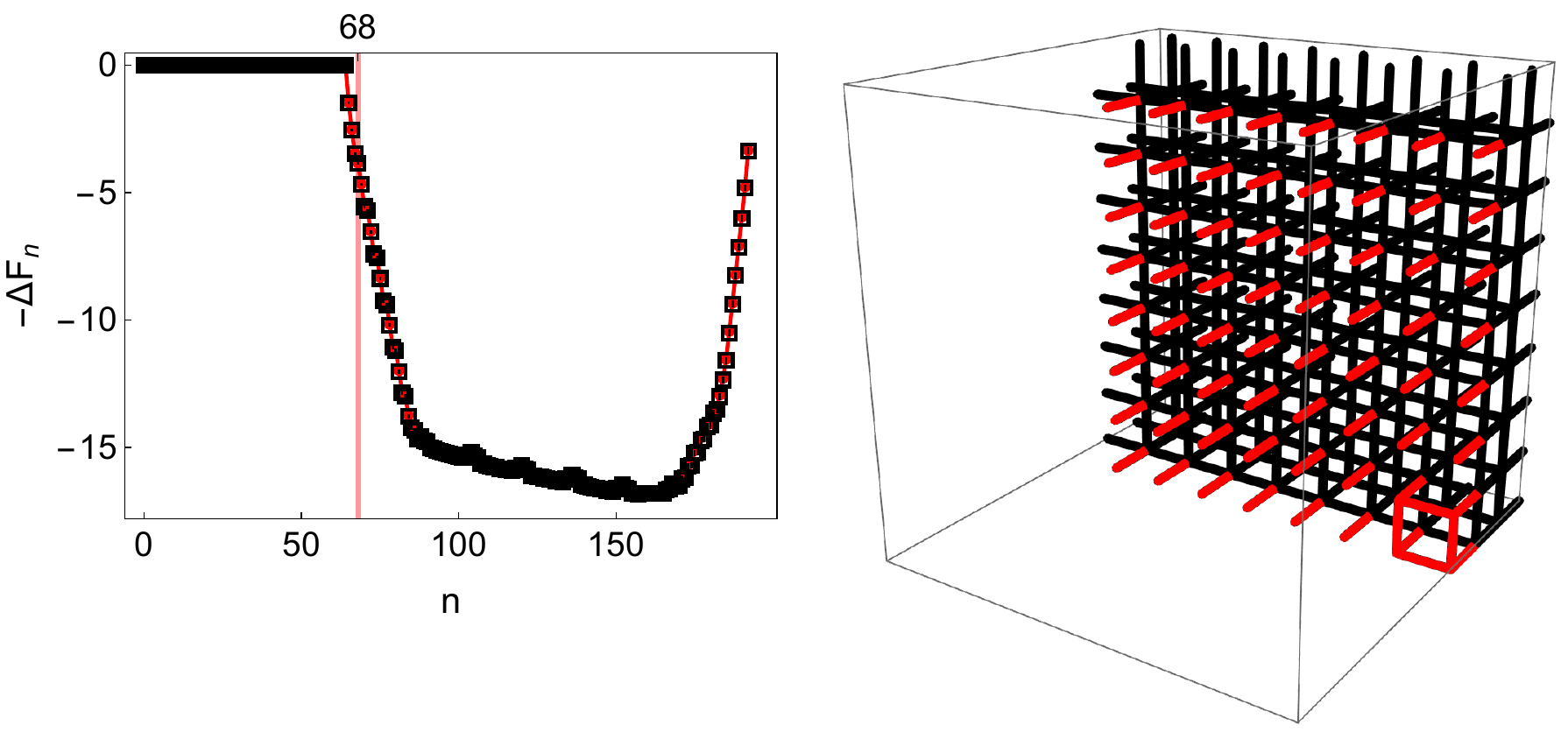}\\[5pt]
\includegraphics[width=\linewidth,keepaspectratio,bgcolor=black!2]{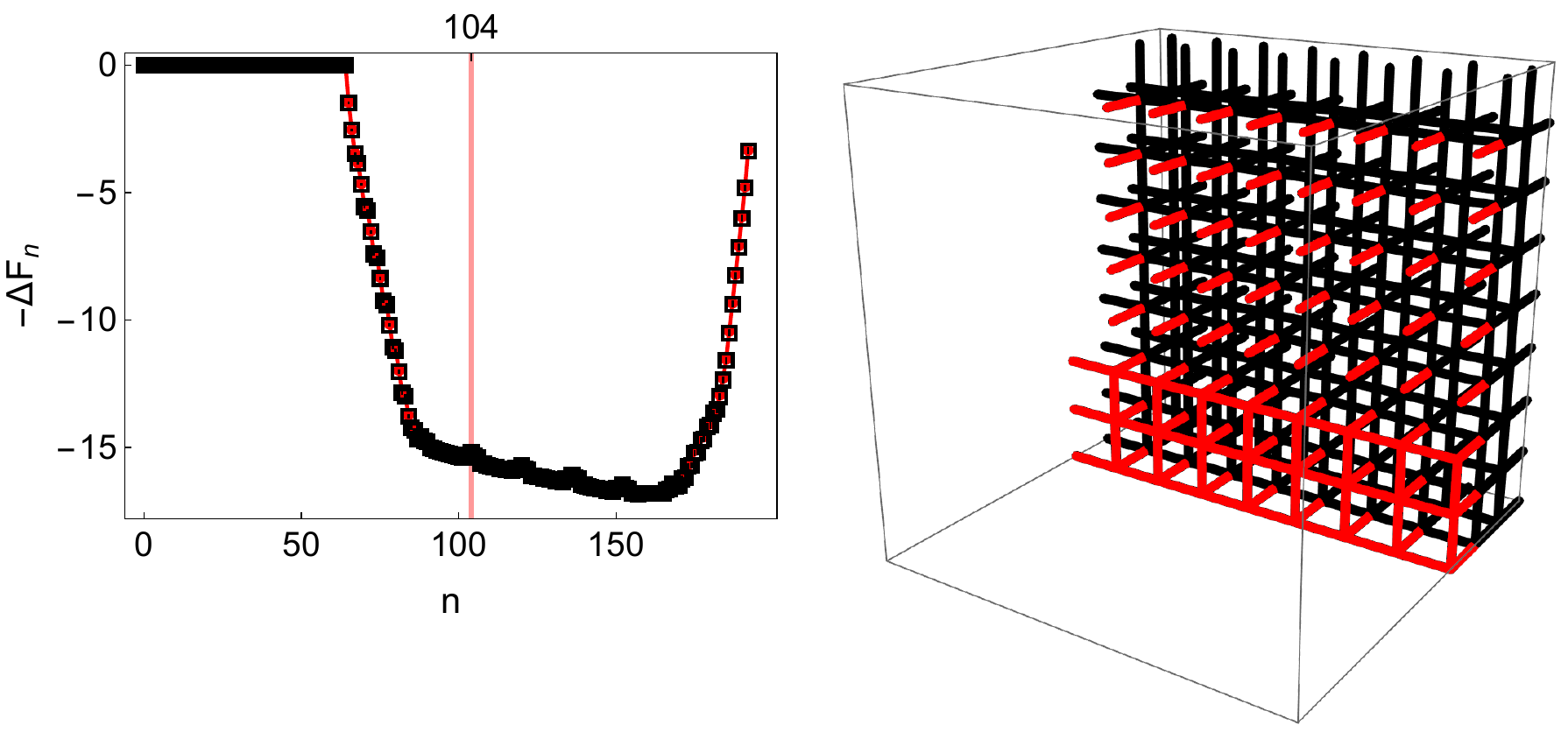}
\end{minipage}\hfill
\begin{minipage}[t]{0.495\linewidth}
\centering
\includegraphics[width=\linewidth,keepaspectratio,bgcolor=black!2]{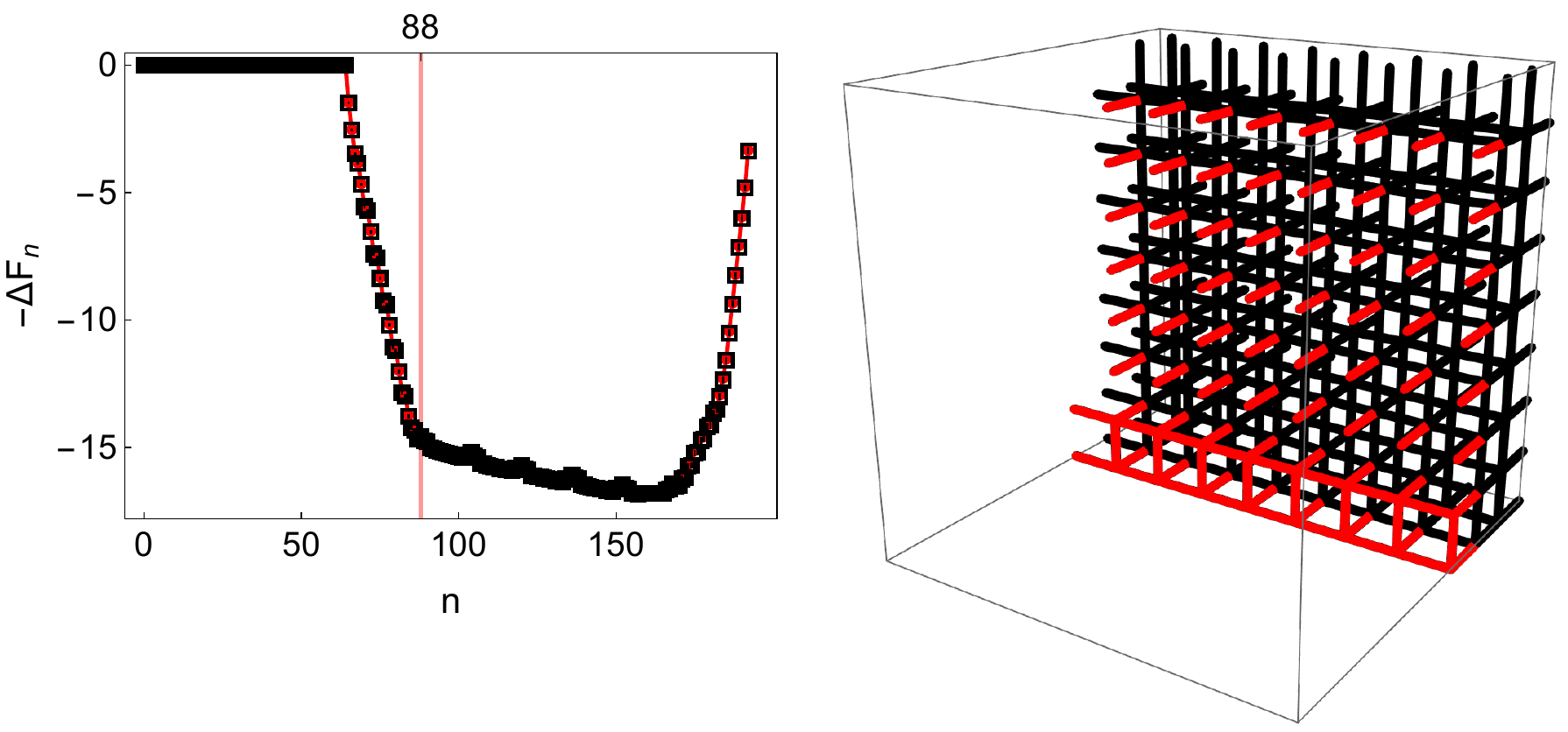}\\[5pt]
\includegraphics[width=\linewidth,keepaspectratio,bgcolor=black!2]{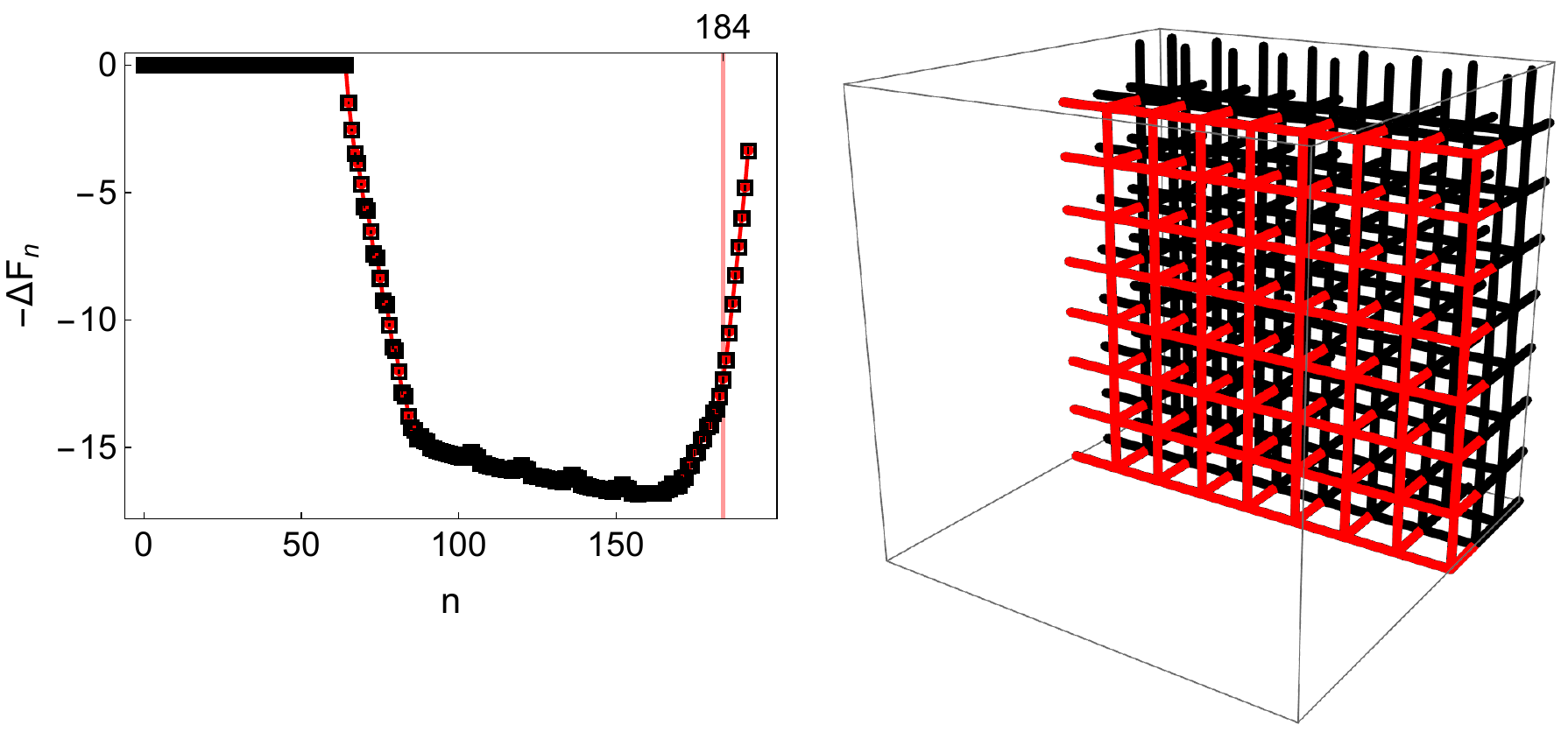}
\end{minipage}
\caption{Same as in Fig.~\ref{fig:freeenergyvsn3d} but for $\SU{3}$ on a (3+1)-dimensional lattice with $N_s=8$, $N_t=16$, $s=2$, and $\beta=6.0$. After all spatial links perpendicular to the boundary of region A have been added, one continues to add spatial links to form complete cubes (cf. top left pair of images) and to fill with them one row after the other. With this ordering, the free-energy difference as function of $n$ is piecewise linear. The here visible free-energy barrier is much smaller than the one that would occur with the method discussed in Sec.~\ref{sec:overlapproblem}.}
  \label{fig:freeenergyvsn4d}
\end{figure}

A possible choice of ordering for the spatial link variables in $K$ is illustrated in Fig.~\ref{fig:freeenergyvsn3d} for the (2+1)-dimensional and in Fig.~\ref{fig:freeenergyvsn4d} for the (3+1)-dimensional case. These orderings use that a change of temporal boundary conditions over spatial links for which one end is always (before and after the update) either completely in region A or completely in region B, does not change the free energy, due to gauge invariance. The free energy graphs in Figs.~\ref{fig:freeenergyvsn3d} and \ref{fig:freeenergyvsn4d} are therefore initially flat and to determine the total free energy difference, Eq.~\eqref{eq:imprfreeenergydiff2}, one has to consider only the boundary states for $n>N_s^{d-2}$. The choice of ordering of the spatial links in $K$ renders the free energy, $\Delta F_n$, as function of $n$ also piecewise linear, which can be used to reduce the computational cost on large systems by determining the initial estimate for $\Delta F_n$ from slope measurements in the linear segments. 

\section{Results}\label{sec:results}
\begin{figure}[h]
\centering
\begin{minipage}[t]{0.44\linewidth}
\centering
\includegraphics[height=0.79\linewidth,keepaspectratio]{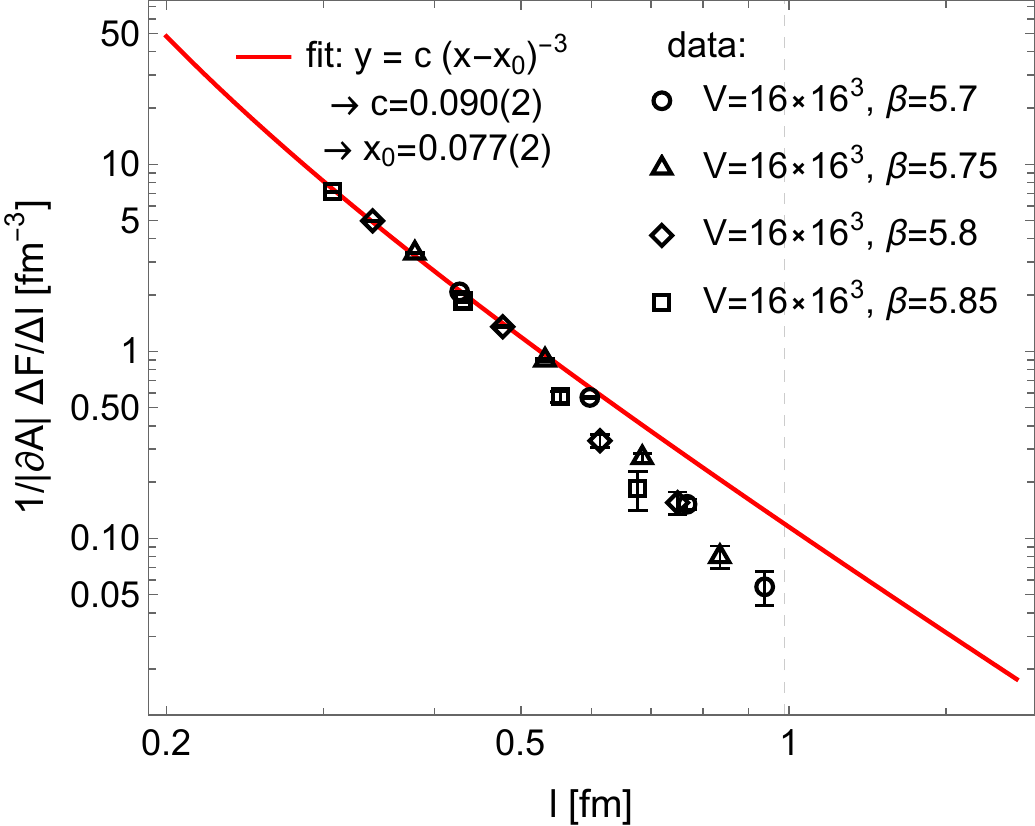}
\caption{Derivative of entanglement entropy from Eq.~\eqref{eq:imprfreeenergydiff} for $\SU{3}$ on a $V=s\,N_t\times N_s^3$ lattice with $N_s=16$, $N_t=16$, $s=2$, plotted as function of $l$ for four different values of $\beta$. The red line represents a fit of the indicated form to the short-distance data ($l<0.5\,\mathrm{fm}$).}
  \label{fig:results1}
\end{minipage}\hfill
\begin{minipage}[t]{0.54\linewidth}
\centering
\includegraphics[height=0.615\linewidth,keepaspectratio]{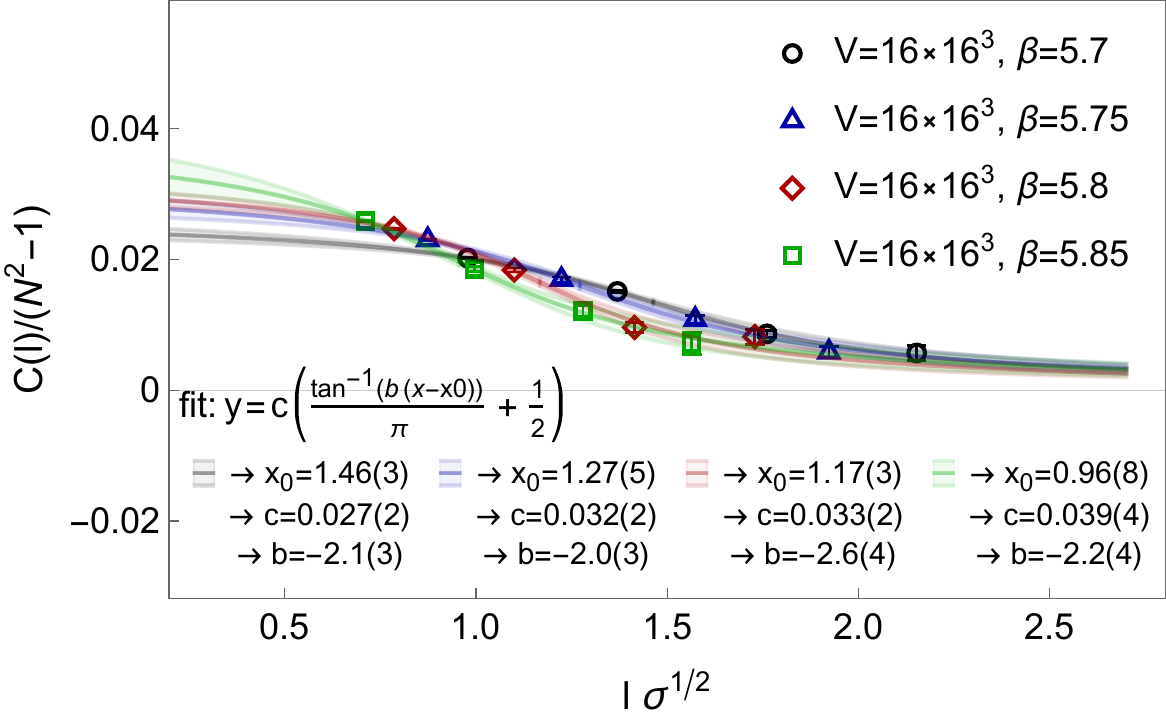}
\caption{Entropic c-function from Eq.~\eqref{eq:cfunc} for the same data as in Fig.~\ref{fig:results1}, but plotted as function of $l\,\sigma^{1/2}$, with $\sigma$ being the string tension. For each beta value the data has been fitted with the indicated function, which has been used also in~\cite{Rabenstein:2018bri}. The data appears to resolve finite $a$ and/or finite volume effects.}
  \label{fig:results2}
\end{minipage}
\end{figure}

We briefly report here some first results obtained with our new method for $\SU{3}$ in (3+1) dimensions. In Fig.~\ref{fig:results1} we show the derivative of the entanglement entropy from Eq.~\eqref{eq:imprfreeenergydiff} as function of the slab-width $l$. The data was obtained on lattices of size $s\,N_t\times N_s^3$ with $s=2$, $N_t=16$ and $N_s=16$ at $\beta\in\cof{5.7,\,5.75,\,5.8,\,5.85}$. We try to fit the data for $l<0.5\,\mathrm{fm}$ to the form $y=c\,\of{x-x_0}^{-3}$ and find $c=0.090(2)$ and $x_0=0.077(2)\,\mathrm{fm}$. The inclusion of the shift $x_0$ is to some extent arbitrary as from holography one expects $x_0=0$, i.e. a pure power law $y\sim x^{-3}$. 

As discussed in~\cite{Rabenstein:2018bri}, a more sensible quantity to consider instead of Eq.~\eqref{eq:imprfreeenergydiff} itself is the the higher-dimensional analogue of the entropic c-function,
\[
C\of{l}=\frac{l^3}{2\,N_s^2}\,\partd{S_{EE}\of{l,N_t,N_s}}{l}\ ,\label{eq:cfunc}
\]
which is expected to be proportional to the number of degrees of freedom (DOF) at scale $l$. If Eq.~\eqref{eq:imprfreeenergydiff} would follow a pure power law $\sim l^{-3}$, then Eq.~\eqref{eq:cfunc} would be constant. The authors of~\cite{Rabenstein:2018bri}, however, showed that their simulation results for Eq.~\eqref{eq:cfunc} can be fitted reasonably well to the form
\[
y=c\of{\frac{\tan^{-1}\of{b\of{x-x_0}}}{\pi}+\frac{1}{2}}\ ,\label{eq:cfuncfitform}
\]
which is qualitativley consistent with a change of DOFs from free gluons deep in the UV to gauge invariant objects (e.g. glueballs) at larger scales. As our data covers approximately the same simulation parameters as the $\SU{3}$ data in~\cite{Rabenstein:2018bri}, also our data represents this transition region where DOFs change, and should therefore indeed not follow a pure power law $\sim l^{-3}$. 

With the reduced errors in our data for Eq.~\eqref{eq:cfunc}, a collective fit of Eq.~\eqref{eq:cfuncfitform} to the data from different $\beta$-values does not appear to work out very well; we therefore show in Fig.~\ref{fig:results2} results from fitting Eq.~\eqref{eq:cfuncfitform} separately to the data for different $\beta$-values, revealing some systematic $\beta$-dependency.

\section{Conclusions and outlook}\label{sec:conclusions}
We have presented a new method for determining R{\'e}nyi entropies of spatial sub-regions in $\SU{N}$ gauge theories with lattice Monte Carlo and performed some first tests with $\SU{3}$ gauge theory in (3+1) dimensions. In (3+1) dimensions, the method is not yet fully optimized, as discussed in the last paragraph of Sec.~\ref{sec:overlapproblemimproved}. However, as illustrated in Sec.~\ref{sec:results}, the current implementation allows already for significantly reduced errors in the measurement of the second order R{\'e}nyi entropy, used as approximation to the entanglement entropy on the lattice. In addition to the here discussed case, we have applied the method also to $\SU{2}$ at finite temperature in (2+1) dimensions and to $\SU{2}$ and $\SU{5}$ in (3+1) dimensions at zero temperature.\footnote{Publications are in preparation.} We are currently working on reducing the computational cost for the (3+1)-dimensional case along the lines mentioned at the end of Sec.~\ref{sec:overlapproblemimproved}.


\begin{thebibliography}{99}
\bibitem{Renyi:1965aa}
A.~R{\'e}nyi,
``On the Foundations of Information Theory,'' 
Revue de l'Institut International de Statistique / Review of the International Statistical Institute Vol. 33, No. 1 (1965), pp. 1-14
doi:\href{https://doi.org/10.2307/1401301}{10.2307/1401301}.

\bibitem{Calabrese:2004eu}
P.~Calabrese and J.~L.~Cardy,
``Entanglement entropy and quantum field theory,''
J. Stat. Mech. \textbf{0406} (2004), P06002
doi:\href{https://doi.org/10.1088/1742-5468/2004/06/P06002}{10.1088/1742-5468/2004/06/P06002}
[arXiv:\href{https://arxiv.org/abs/hep-th/0405152}{hep-th/0405152} [hep-th]].

\bibitem{Rabenstein:2018bri}
A.~Rabenstein, N.~Bodendorfer, P.~Buividovich and A.~Sch\"afer,
``Lattice study of R\'enyi entanglement entropy in $SU(N_c)$ lattice Yang-Mills theory with $N_c = 2, 3, 4$,''
Phys. Rev. D \textbf{100} (2019) no.3, 034504
doi:\href{https://doi.org/10.1103/PhysRevD.100.034504}{10.1103/PhysRevD.100.034504}
[arXiv:\href{https://arxiv.org/abs/1812.04279}{1812.04279} [hep-lat]].

\bibitem{Buividovich:2008kq}
P.~V.~Buividovich and M.~I.~Polikarpov,
``Numerical study of entanglement entropy in SU(2) lattice gauge theory,''
Nucl. Phys. B \textbf{802} (2008), 458-474
doi:\href{https://doi.org/10.1016/j.nuclphysb.2008.04.024}{10.1016/j.nuclphysb.2008.04.024}
[arXiv:\href{https://arxiv.org/abs/0802.4247}{0802.4247} [hep-lat]].

\bibitem{Nakagawa:2009jk}
Y.~Nakagawa, A.~Nakamura, S.~Motoki and V.~I.~Zakharov,
``Entanglement entropy of SU(3) Yang-Mills theory,''
PoS \textbf{LAT2009} (2009), 188
doi:\href{https://doi.org/10.22323/1.091.0188}{10.22323/1.091.0188}
[arXiv:\href{https://arxiv.org/abs/0911.2596}{0911.2596} [hep-lat]].

\bibitem{Wang:2000fzi}
F.~Wang and D.~P.~Landau,
``Efficient, Multiple-Range Random Walk Algorithm to Calculate the Density of States,''
Phys. Rev. Lett. \textbf{86} (2001) no.10, 2050
doi:\href{https://doi.org/10.1103/PhysRevLett.86.2050}{10.1103/PhysRevLett.86.2050}
[arXiv:\href{https://arxiv.org/abs/cond-mat/0011174}{cond-mat/0011174} [cond-mat.stat-mech]].

\end{thebibliography}
\end{document}